\documentclass[a4paper,fleqn,usenatbib]{mnras}

\usepackage[T1]{fontenc}
\usepackage{ae,aecompl}

\usepackage{graphicx}	
\usepackage{amsmath}	
\usepackage{amssymb}	

\title[SPE-induced radiation dose]{Modeling Stellar Proton Event-induced particle radiation dose on close-in exoplanets}

\author[Dimitra Atri]{
Dimitra Atri,$^{1}$\thanks{E-mail: dimitra@bmsis.org}
\\
$^{1}$Blue Marble Space Institute of Science, 1001 4th Ave, Suite 3201, Seattle, WA 98154\\
}

\date{Accepted XXX. Received YYY; in original form ZZZ}

\pubyear{2016}

\begin{document}
\label{firstpage}
\pagerange{\pageref{firstpage}--\pageref{lastpage}}
\maketitle

\begin{abstract}
{\it Kepler} observations have uncovered the existence of a large number of close-in exoplanets and serendipitously of stellar superflares with emissions several orders of magnitude higher than those observed on the Sun. The interaction between the two and its implications on planetary habitability is of great interest to the community. Stellar Proton Events interact with planetary atmospheres, generate secondary particles and increase the radiation dose on the surface. This effect is amplified for close-in exoplanets and can be a serious threat to potential planetary life. Monte Carlo simulations are used to model the SPE-induced particle radiation dose on the surface of such exoplanets. The results show a wide range of surface radiation doses on planets in close-in configurations with varying atmospheric column depths, magnetic moments and orbital radii. It can be concluded that for close-in exoplanets with sizable atmospheres and magnetospheres, the radiation dose contributed by stellar superflares may not be high enough to sterilize a planet (for life as we know it) but can result in frequent extinction level events. In light of recent reports, the interaction of hard-spectrum SPEs with the atmosphere of Proxima Centauri b is modeled and implications on its habitability is discussed.
\end{abstract}




\section{Introduction}

{\it Kepler} observations have revealed the existence of a large number of exoplanets orbiting in close-in configurations around a variety of stars. Due to proximity to the host star ($\sim$ 0.01 AU in some cases), such planets are greatly influenced by its activity, such as stellar wind, and abrupt emissions in form of Coronal Mass Ejections (CMEs) and flares. Flares are often accompanied by bursts of energetic protons, also known as Stellar Proton Events (SPEs). SPEs, CMEs and stellar wind directly interact with the exoplanet's atmosphere and are capable of enhancing atmospheric depletion \citep{2015MNRAS.449.4117V} and photochemical reaction rates \citep{segura2005biosignatures,segura2010effect}. These effects are especially relevant in case of M dwarfs, whose habitable zones are typically $\sim$ 0.1 AU from the host star with high flaring activity. M dwarfs are faint, low mass stars, with extremely long main sequence lifetimes, and represent about 70\% of the total population of stars in the Milky Way. It has been shown that only 100 g\,cm$^{-2}$ of CO$_2$ is required to support liquid water on the surface of tidally-locked close-in planets around such stars \citep{haberle1996can}. A lot of effort has been made to model the effects of flares on climate and photochemistry of earth-like planets in close-in orbits \citep{segura2005biosignatures,segura2010effect,2016A&A...585A..96T}. These studies of atmospheric changes resulting from stellar flares take into account the X-ray, EUV and proton flux from the host star. Because of their low energy (keV-MeV), stellar wind particles are not energetic enough to penetrate below the upper atmosphere, but ``hard" stellar protons (GeV) do have the capability.

There has been no study where the direct impact of SPE-induced charged particles on the exoplanet surface has been considered. The aim of this manuscript is to go beyond the impact of stellar flares on the atmosphere and model the production and propagation of secondary particles generated by SPEs interacting with the exoplanet atmosphere and calculate the radiation dose on its surface, and to better understand the role of planetary atmospheric size (column density), magnetic field and orbital distance from the host star on surface radiation dose and its implications on planetary habitability. 

Protons with energies greater than the pion production threshold (290 MeV) produce secondary particles and initiate a cascade of particles with sufficiently energetic ones capable reaching the planet's surface. On Earth, such events are called Ground Level Enhancements (GLEs) and for exoplanets the equivalent term would be Surface Level Enhancements (SLEs). These high-energy stellar particles with energies typically up to 10 GeV, contribute both to the atmospheric ionization and radiation dose on the surface and significant doses can potentially have biological implications \citep{ferrari2009cosmic,atri2010lookup,dartnell2011ionizing,melott2011astrophysical,atri2013galactic,atri2011modeling,atri2014cosmic,2016A&A...585A..96T}. The flux of stellar particles increases by several orders of magnitude for timescales of hours during intense SPEs, and become more important in case of close-in exoplanets. Hard-spectrum events are a subset of all SPEs emitted by a star. Soft-spectrum SPEs have significant effects in the upper atmosphere and do not produce GLEs, whereas hard-spectrum events produce GLEs with less effect in the upper atmosphere. Galactic Cosmic Rays (GCRs), which are of much higher energy but of lower flux, also contribute to the radiation dose on the surface of such planets \citep{atri2013galactic,griessmeier2016galactic}. 

Even though there are numerous observations of stellar flares, the details of particle spectrum is difficult to determine and the best approach is to model the effect of such events based on well-studied solar events. Flares with fluence (at 1 AU) $<$ 10$^9$ protons\,cm$^{-2}$ ($\sim$ 10$^{31}$ ergs) are recorded only a very few times per solar cycle on Earth \citep{smart2006}. Events with higher fluence are less frequent, with the Carrington event (1859) with a fluence (at 1 AU) of $\sim$10$^{10}$ protons\,cm$^{-2}$ ($\sim$ 10$^{32}$ ergs) occurring at a rate of about once a century. The AD 775 event had an estimated energy of $\sim$ 10$^{33}$ ergs. {\it Kepler} survey has observed a large number of energetic flares, or superflares of energies 10$^{33}$ to 10$^{36}$ ergs around other stars \citep{maehara2012superflares,shibayama2013,candelaresi2014superflare}. Studies have also shown that superflares on M stars are 10-100 times more frequent than on G stars \citep{maehara2012superflares}. It is the goal of the manuscript to quantify the surface radiation dose from hard-spectrum events on close-in exoplanets and discuss its implications on constraining planetary habitability. 
\section{Numerical Modeling}
The radiation dose on the surface of the planet is governed by the flux and energy of incident particles, atmospheric depth and the strength of its magnetosphere. These effects are studied here by choosing well-studied SPEs with extensive GLE measurements on Earth and modeling its interactions with planetary atmospheres. Three such high fluence events occurred on 23 February 1956 (SPE56), 4 August 1972 (SPE72) and 29 September 1989 (SPE89). The solar event spectra are represented by Band functions giving the event integrated fluence for protons with energies between 10 MeV - 10 GeV \citep{tylka2009new}. Figure 1 shows the spectra of the three events. The events SPE56 and SPE72 represent the ``hard" and ``soft" spectra of large fluence events respectively. The solar event SPE89 falls between the two, and for reasons discussed later will be the focus of this work. Since it occurred in the late 80s, it is also one of the most well-studied high-fluence events accompanied with a GLE with both satellite and ground-based observations. 

Since incident charged particle flux strongly depends on the planetary magnetic field, the next step was to calculate the spectra of penetrating particles for different magnetic field configurations. In order to accomplish this, the magnetospheric filter function to calculate the energy dependent shielding efficiency of planets with differing magnetic moments was needed. The magnetospheric filter function gives the efficiency of particles penetrating the atmosphere and is defined as the ratio of the number of shielded and unshielded particles (n$_{shielded}$/n$_{unshielded}$) \citep{griessmeier2015galactic}. The results of simulations from our earlier work were used and details of all the calculations can be found in the cited manuscript \citep{griessmeier2015galactic}. Figure 2 is based on results from \citep{griessmeier2015galactic}, and shows the efficiency of penetration of protons with increasing energies (10 MeV to 32 GeV) on planets with different magnetic field strengths (0.05 ${\cal M}$$_{Earth}$ to 10 ${\cal M}$$_{Earth}$) where 1 ${\cal M}$$_{Earth}$=7.94$\times$10$^{22}$ Am$^2$. This filter function was used to calculate the input spectrum in each scenario for atmospheric modeling. 

GEANT4 is a widely used Monte Carlo package that models the propagation of charged particles in planetary atmospheres and performs a variety of calculations such as computing radiation dose \citep{agostinelli2003geant4}. The code simulates particle interactions and tracks particles down to the surface level defined by the user. In order to model SPE-induced radiation dose, simulations using the model with 6 atmospheric sizes of 30, 70, 100, 300, 700 and 1000 g\,cm$^{-2}$ were performed and radiation doses at the surface level were computed. The earth's present column density is 1036$\simeq$1000 g\,cm$^{-2}$. The atmospheric interaction of SEPs was simulated by obtaining the input spectra using Band functions, by applying the magnetospheric filter functions described above and simulating 10$^9$ protons for each event from 10 MeV to 10 GeV, incident from different angles over the hemisphere. The magnetic field was switched off since the cutoff rigidities were taken from the magnetospheric filter functions. Although, the Earth's atmospheric composition was used for this work, it must be emphasized that the calculations done here depend primarily on column density (g\,cm$^{-2}$) and are weakly dependent on the composition. The interactions modeled here depend on the number of nucleons per g\,cm$^{-2}$, and the numbers are very similar for typical C, N, O atmospheres (12-16 g mol$^{-1}$). The calculations might change for a pure hydrogen atmosphere but that case will not be considered here. As explained later, the same event was used and further rescaled for the total energy between 10$^{32}$ - 10$^{36}$ ergs to estimate the possible effects from a wide range of events such as superflares. These events were also rescaled to orbital distances of 0.01-0.2 AU using the r$^2$ scaling factor since fluence $\sim$ ($\theta$$_{Opening}$\,R$_{Orbit}^2$)$^{-1}$. It must be noted that since these events occur on timescales of hours, the GCR-induced dose during the period is much smaller than the high-fluence event-induced dose which is the focus of this work; and is likely going to decrease for close-in planets due to stellar wind shielding, and therefore will be ignored \citep{atri2013galactic}.

\begin{figure}
\centering 
\includegraphics[width=0.47\textwidth]{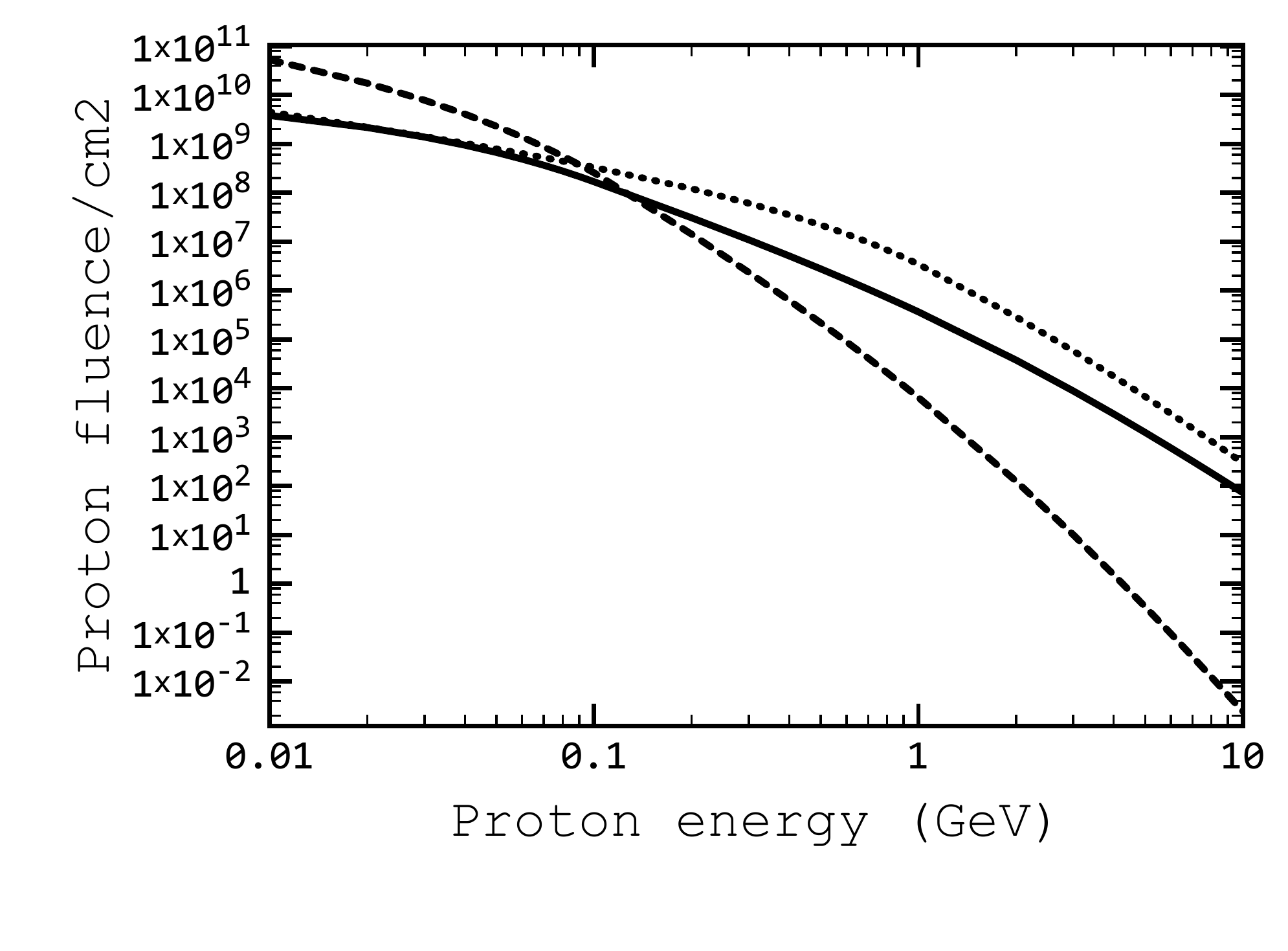}
\caption{\label{fig:i} Event integrated spectra of the 4 August 1972 (dash), 29 September 1989 (solid) and 23 February 1956 (dots) solar protons events on Earth, based on parameters obtained from \citep{tylka2009new}.}
\end{figure}

\begin{figure}
\centering 
\includegraphics[width=0.47\textwidth]{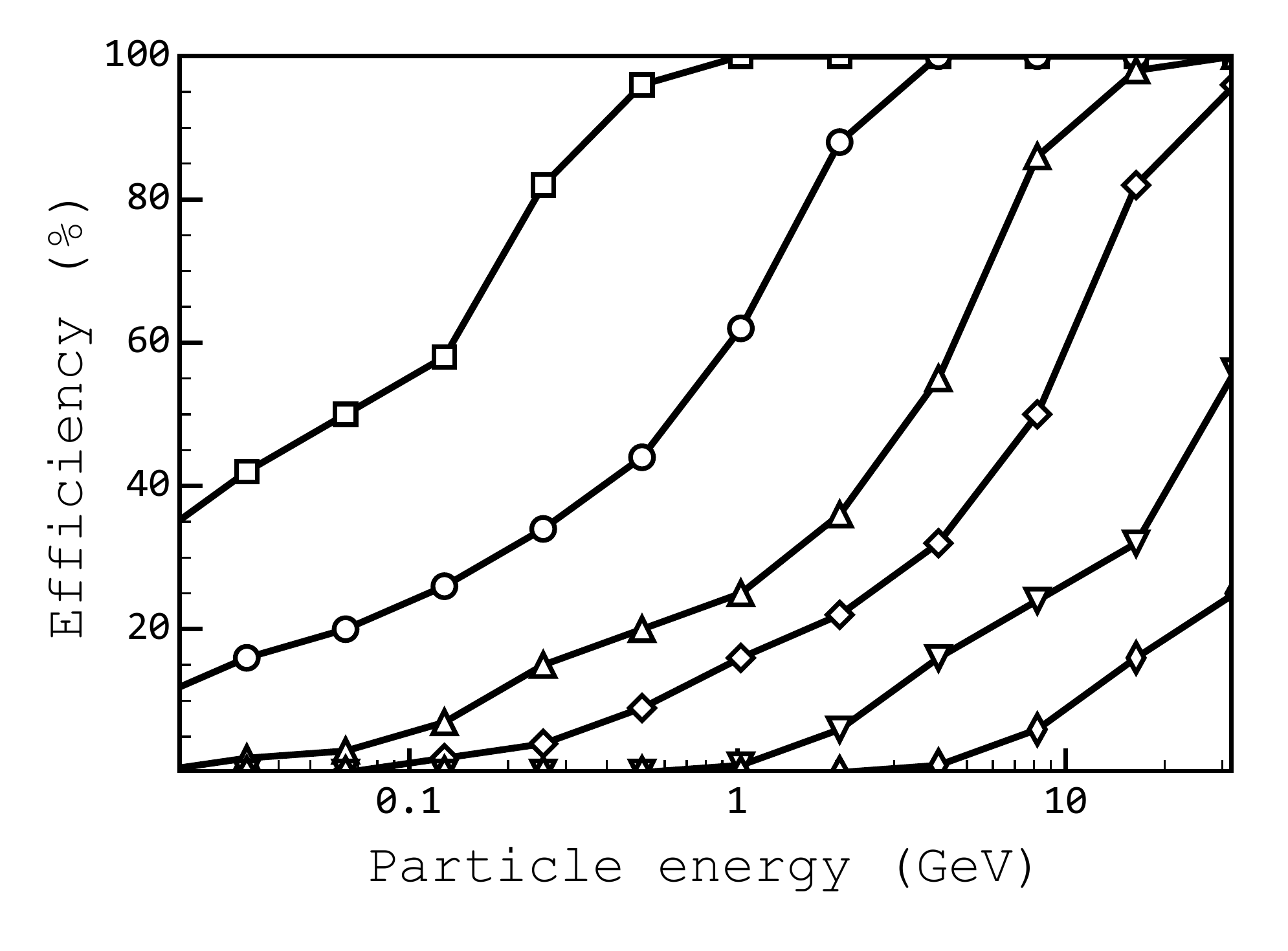}
\caption{\label{fig:i} Efficiency of energetic protons penetrating with magnetospheric filter function obtained from \citep{griessmeier2015galactic}. 0.05 ${\cal M}$$_{Earth}$ (same as 0) on the left represented by the squares, followed by 0.15, 0.5, 1.0, 3.0 to 10 ${\cal M}$$_{Earth}$ all the way to the right represented by diamonds. All values of magnetic moment shown in the figure are relative to the Earth's magnetic moment.}
\end{figure}

\begin{figure}
\centering 
\includegraphics[width=0.47\textwidth]{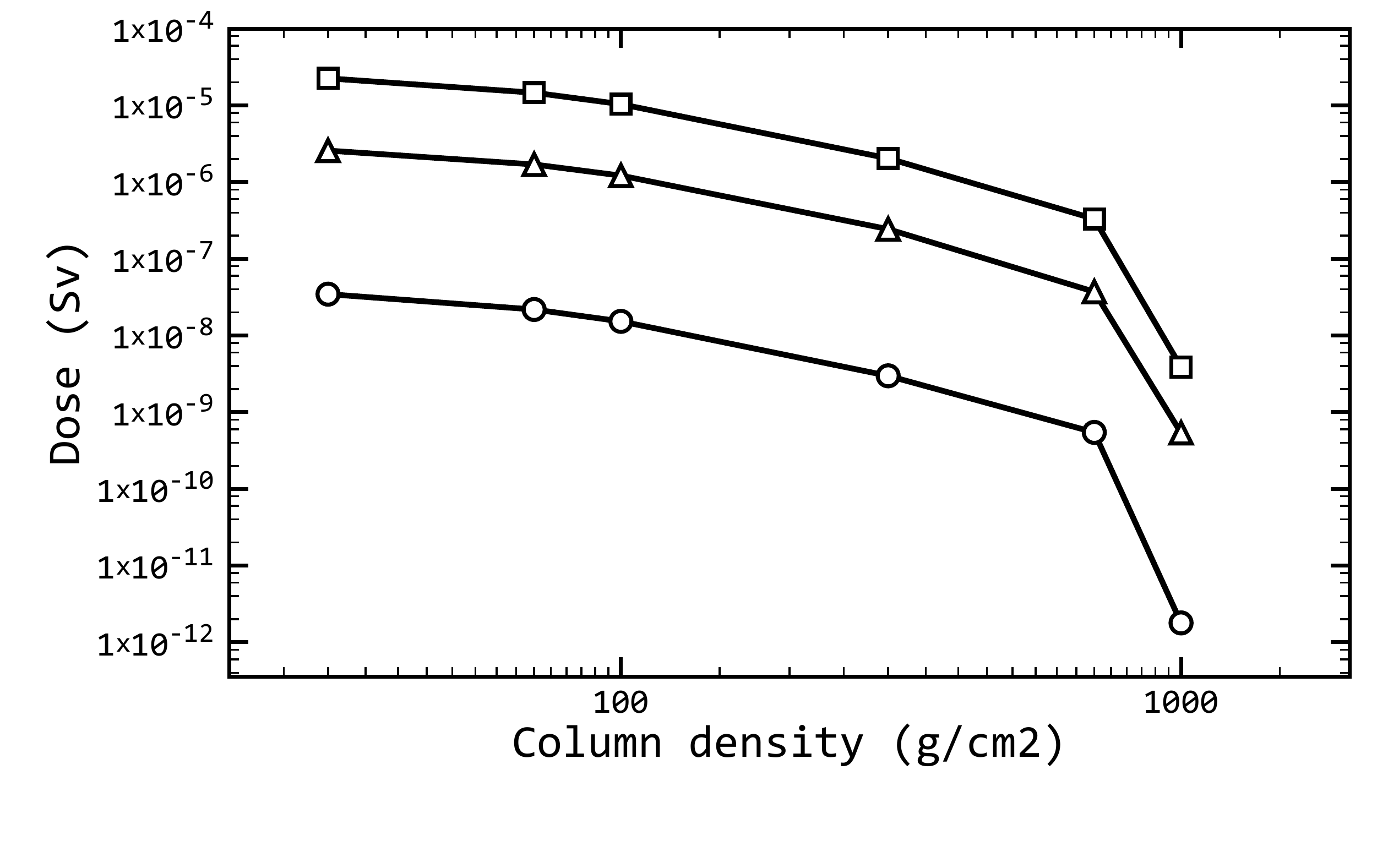}
\caption{\label{fig:i} Radiation dose values from SPE56 (square), SPE89 (triangle) and SPE72 (circle) for various cases of atmospheric column depths at 1 AU with 1 ${\cal M}$$_{Earth}$ magnetic moment.}
\end{figure}

\begin{figure}
\centering 
\includegraphics[width=0.47\textwidth]{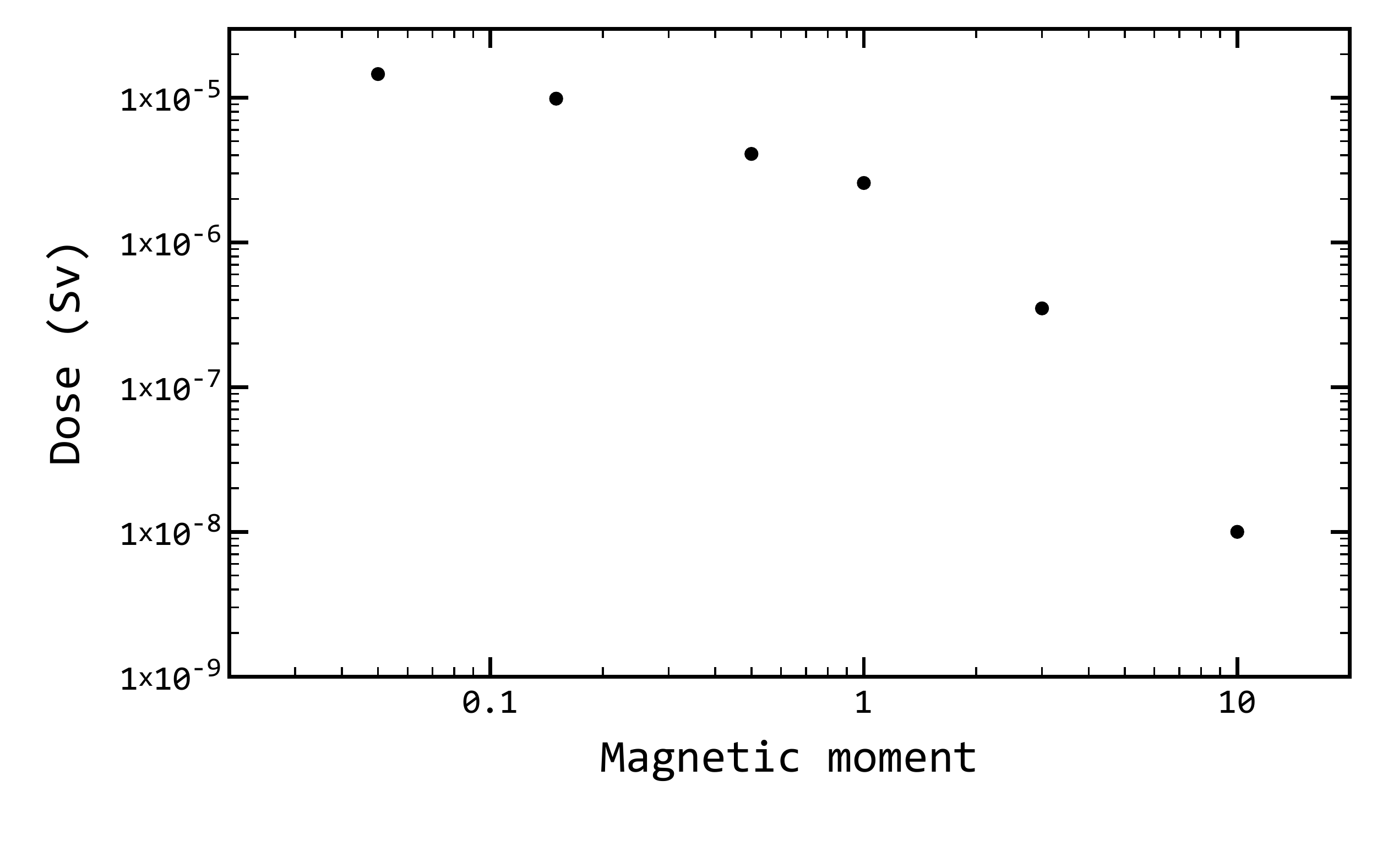}
\caption{\label{fig:i} Radiation dose values as a function of changing planetary magnetic moment from SPE89 with 1000 g\,cm$^{-2}$ atmosphere.}
\end{figure}

\begin{figure}
\centering 
\includegraphics[width=0.47\textwidth]{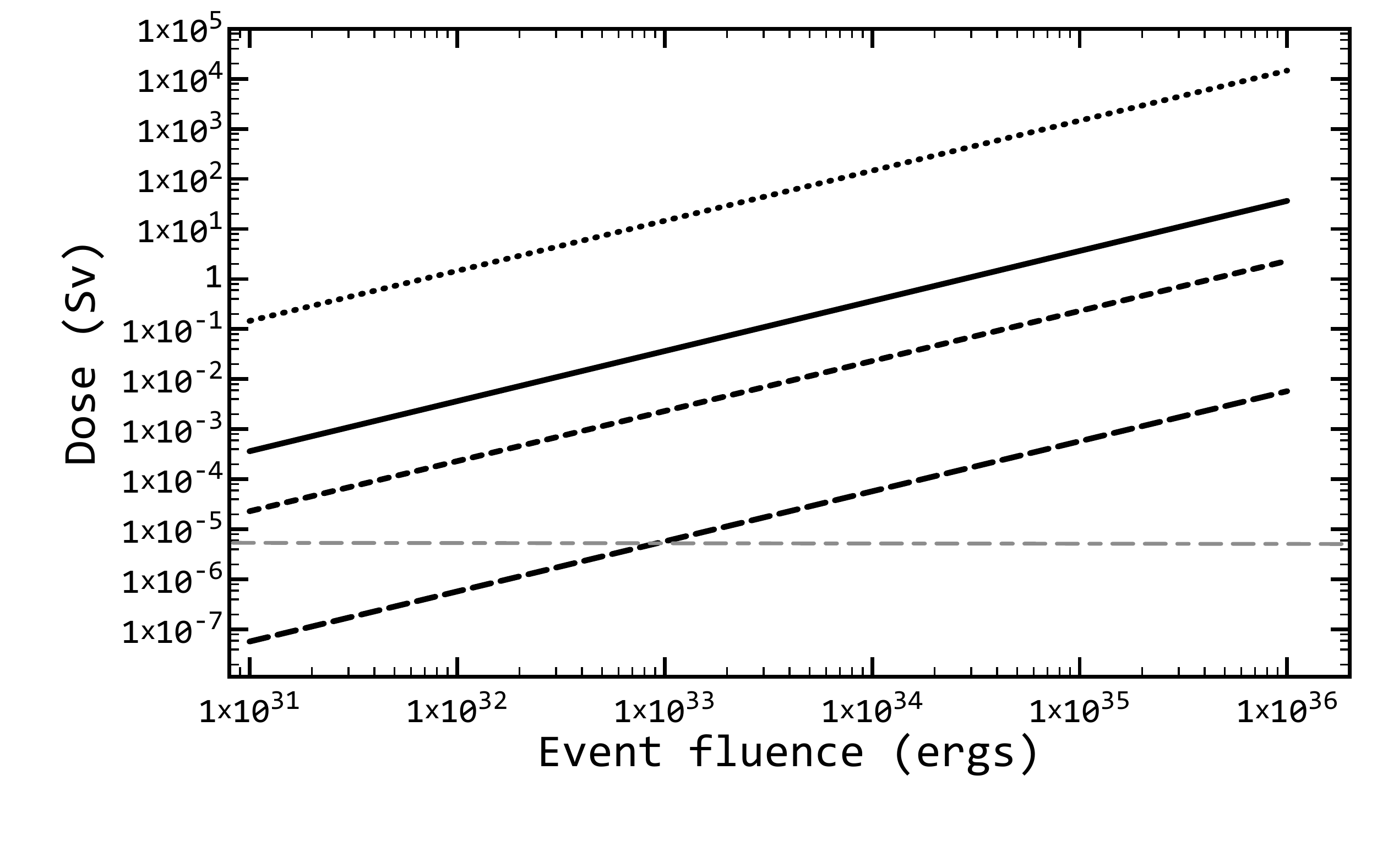}
\caption{\label{fig:i} Radiation dose from SPEs at depths of 30 (dots) and 1000 (short dash) g\,cm$^{-2}$ at 0.01 AU,  30 (solid) and 1000 (long dash) g\,cm$^{-2}$ at 0.2 AU with 0.05 ${\cal M}$$_{Earth}$ magnetic moment. The globally averaged daily dose from natural background radiation on Earth is shown by the horizontal dashed line for comparison.}
\end{figure}

\begin{figure}
\centering 
\includegraphics[width=0.47\textwidth]{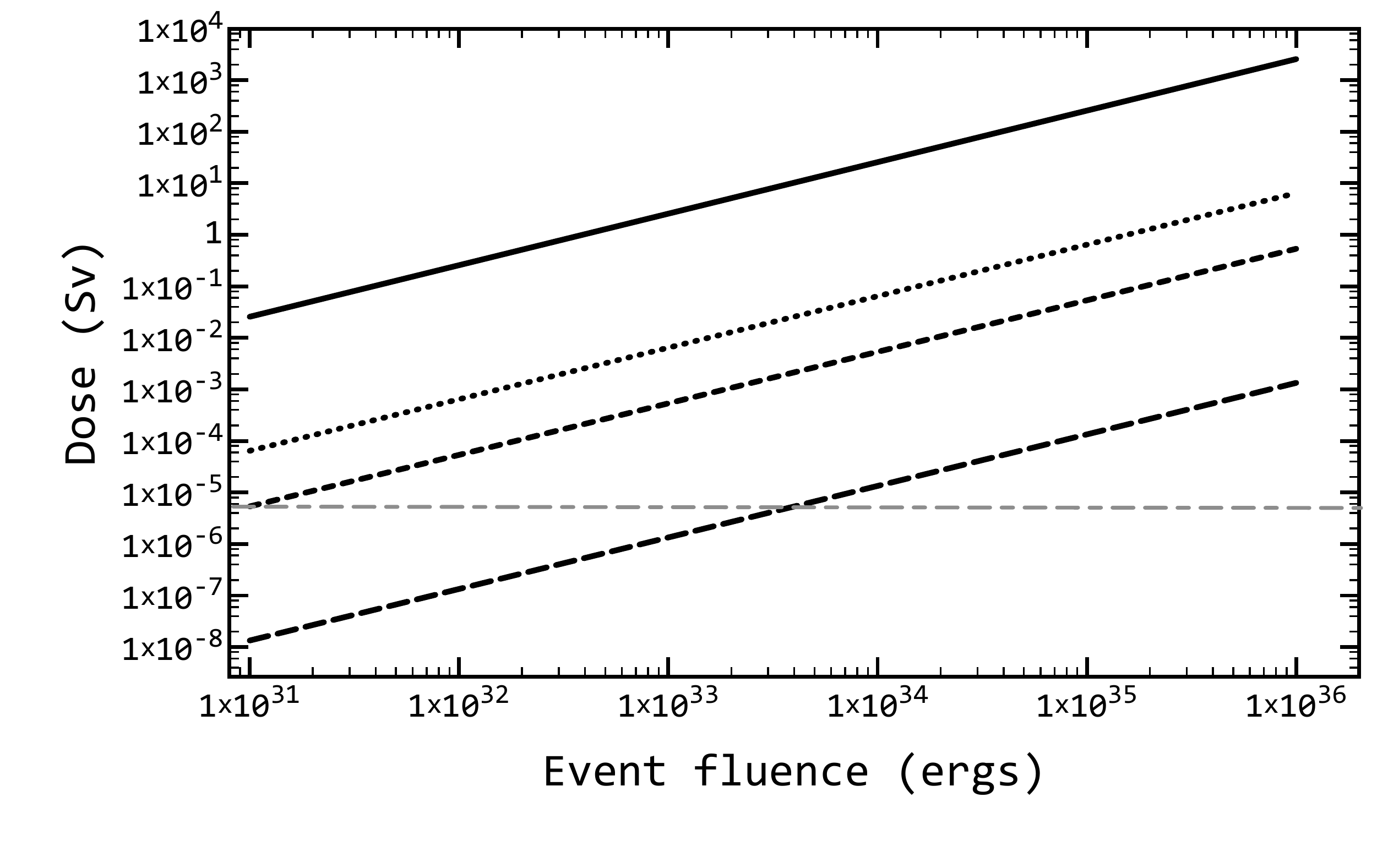}
\caption{\label{fig:i} Radiation dose from SPEs at depths of 30 (solid) and 1000 (short dash) g\,cm$^{-2}$ at 0.01 AU, 30 (dots) and 1000 (long dash) g\,cm$^{-2}$ at 0.2 AU with 1 ${\cal M}$$_{Earth}$ magnetic moment. The globally averaged daily dose from natural background radiation on Earth is shown by the horizontal dashed line for comparison.}
\end{figure}

\begin{figure}
\centering 
\includegraphics[width=0.47\textwidth]{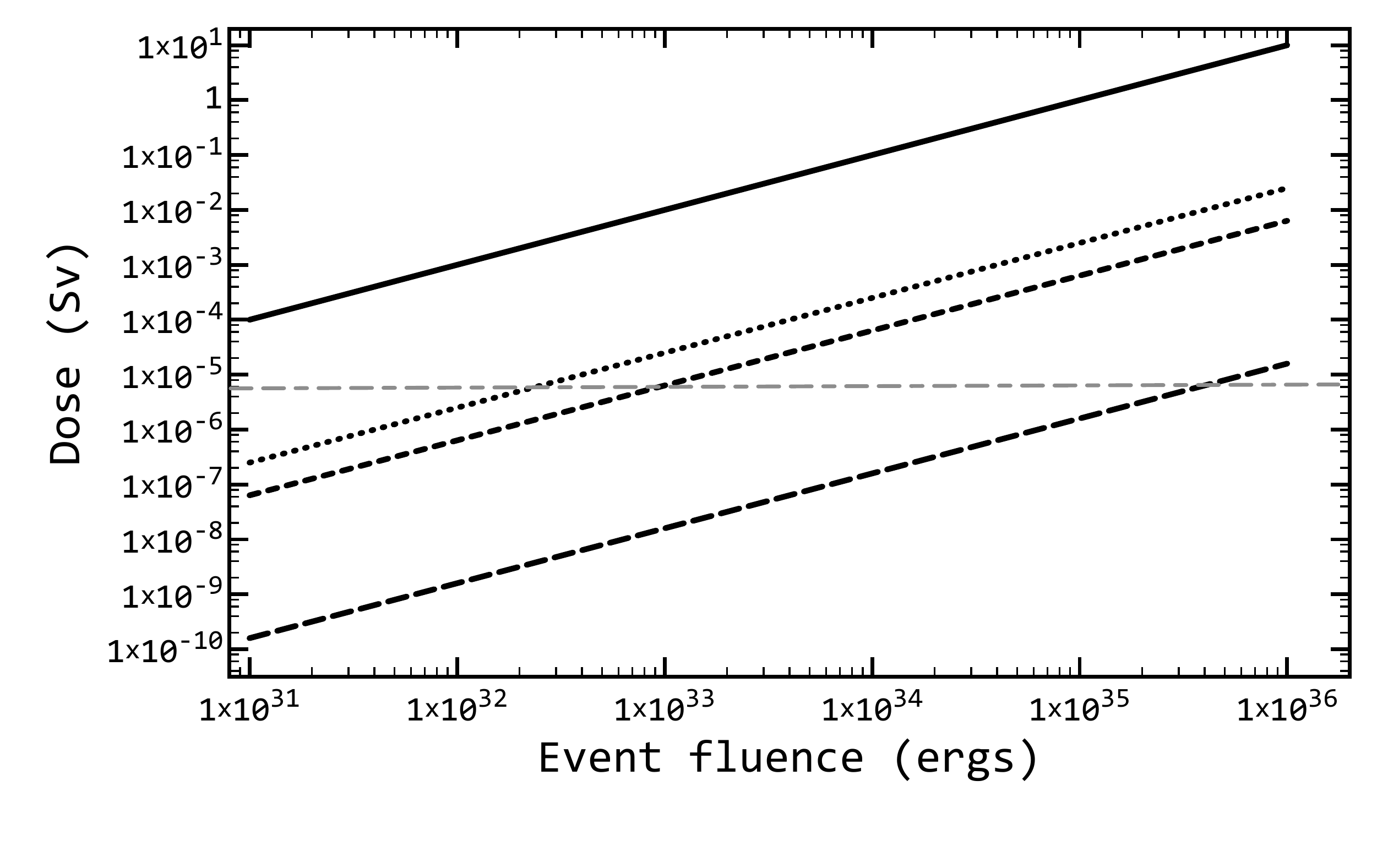}
\caption{\label{fig:i}  Radiation dose from SPEs at depths of 30 (solid) and 1000 (short dash) g\,cm$^{-2}$ at 0.01 AU, 30 (dots) and 1000 (long dash) g\,cm$^{-2}$ at 0.2 AU with 10 ${\cal M}$$_{Earth}$ magnetic moment. The globally averaged daily dose from natural background radiation on Earth is shown by the horizontal dashed line for comparison.}
\end{figure}

\section{Results}
Figure 3 shows the radiation dose values from SPEs 56, 72 and 89 for different cases of atmospheric column depths (30-1000 g\,cm$^{-2}$) with 1 ${\cal M}$$_{Earth}$ magnetic moment and 1 AU orbital distance. As anticipated, the dose values drop significantly with increasing column depths for all three events. All three events displayed here are considered large events, but have significantly different dose depositions. This is because the dose depends on both the event fluence as well as on the spectral ``hardness". SPE72 has the highest fluence amongst the three events but because of a ``soft" spectrum, the deposited dose is the smallest. SPE56 on the other hand has the largest dose amongst the three because of a ``hard" spectrum. SPE89 was therefore chosen for all calculations because it lies between the two extremes and would be a better representation of large events in general. In addition to the spectrum, column depth and fluence, another important factor governing the radiation dose is the planetary magnetic moment. Figure 4 shows the variation of radiation dose from SPE89 with different planetary magnetic moments. There is about three orders of magnitude difference between the radiation dose from a 0.05 ${\cal M}$$_{Earth}$ to 10 ${\cal M}$$_{Earth}$ planet. The estimated event integrated proton fluence of SPE89 ($>$30 MeV) F$_{30}$ is 1.4$\times$10$^9$ protons\,cm$^{-2}$ \citep{smart2006}. The largest recorded solar event in history was the 1859 Carrington event with an estimated F$_{30}$ of 1.1$\times$10$^{10}$ protons\,cm$^{-2}$. There have been observations of flares on stars with several orders of magnitude higher fluence than those observed on the Sun. Segura et al. \citep{segura2010effect} simulated the atmospheric effects of a SPE with a fluence of 1.5$\times$10$^{12}$ protons\,cm$^{-2}$. The the relationship between the event fluence and total energy (in ergs) depends on a number of factors, for example the spectral hardness, opening angle of the event and the fraction of energy going into accelerating particles. In order to scale the total event energy with F$_{30}$, an assumption for F$_{30}$ of $\sim$10$^9$ protons\,cm$^{-2}$ is made, and the total event energy is 10$^{31}$ ergs at 1 AU. This assumption also makes sense since the $\sim$10$^{10}$ protons\,cm$^{-2}$ event described above had an estimated total energy of $\sim$ 10$^{32}$ ergs at 1 AU. This relation will be used to scale the events up to 10$^{36}$ ergs.

Figures 5, 6 and 7 show the radiation dose at depths of 30 and 1000 g\,cm$^{-2}$, at 0.2 and 0.01 AU for planets with magnetic moments of 0.05, 1 and 10 ${\cal M}$$_{Earth}$ respectively. The case with 0.05 ${\cal M}$$_{Earth}$ is a planet with virtually no magnetic field and the radiation dose varies between 1.46$\times$10$^4$ Sv to 5.77$\times$10$^{-8}$ Sv. For comparison, the globally averaged annual dose from natural background radiation on Earth is 2.4 mSv \citep{atri2014cosmic} which equates to 6.6$\times$10$^{-6}$ Sv/day. The most extreme dose, as expected, is for a non magnetized planet with a thin atmosphere. This case is important also because most non magnetized planets will eventually lose their atmospheres due to the impact of stellar wind. This effect would be amplified in close-in scenarios. Figure 6 shows the radiation dose for a planet with 1 ${\cal M}$$_{Earth}$ and the dose lies between 2.57$\times$10$^3$ and 1.34$\times$10$^{-8}$ Sv. The individual doses fall with increasing magnetic shielding and this effect is most amplified with 10 ${\cal M}$$_{Earth}$ magnetic moment where the dose varies between 10 and 1.59$\times$10$^{-5}$ Sv.

\subsection{Radiation dose on Proxima Centauri b}
Proxima Centauri b is a recently discovered exoplanet orbiting our stellar neighbor Proxima Centauri (1.3 parsec), an M dwarf \citep{anglada2016terrestrial}. Even though its atmosphere is yet to be characterized, preliminary analysis suggests that because of its proximity to the host star ($\sim$0.05 AU) it receives about 65\% of the energy that earth receives from the Sun, and the planet might be ``habitable" with reasonable assumptions about its atmosphere. But would a potential ecosystem on the planet be able to survive hard-spectrum superflares? SPE56 is a good proxy for such events as discussed above and was applied to possible atmospheric column densities on Proxima b assuming earth-equivalent planetary magnetic field. Figure 8 shows that if it has a 1000 g\,cm$^{-2}$ atmosphere (like the Earth) and 1 ${\cal M}$$_{Earth}$ magnetic moment, the particle radiation dose from even the most extreme SPE would not have any significant impact on its biosphere. However, for thinner atmospheres (700 g\,cm$^{-2}$ or lower, as shown in figure) it would be able to produce ``extinction level" doses (5-10 Sv) although not enough to sterilize the planet of life as we know it ($\sim$10$^5$ Sv) \citep{harrison1996taxonomic}. The dose would also increase considerably if the magnetic moment is much lower in case the planet is tidally locked \citep{griessmeier2015galactic}. A detailed analysis on Proxima b will be reported in a forthcoming manuscript by the author.

\begin{figure}
\centering 
\includegraphics[width=0.47\textwidth]{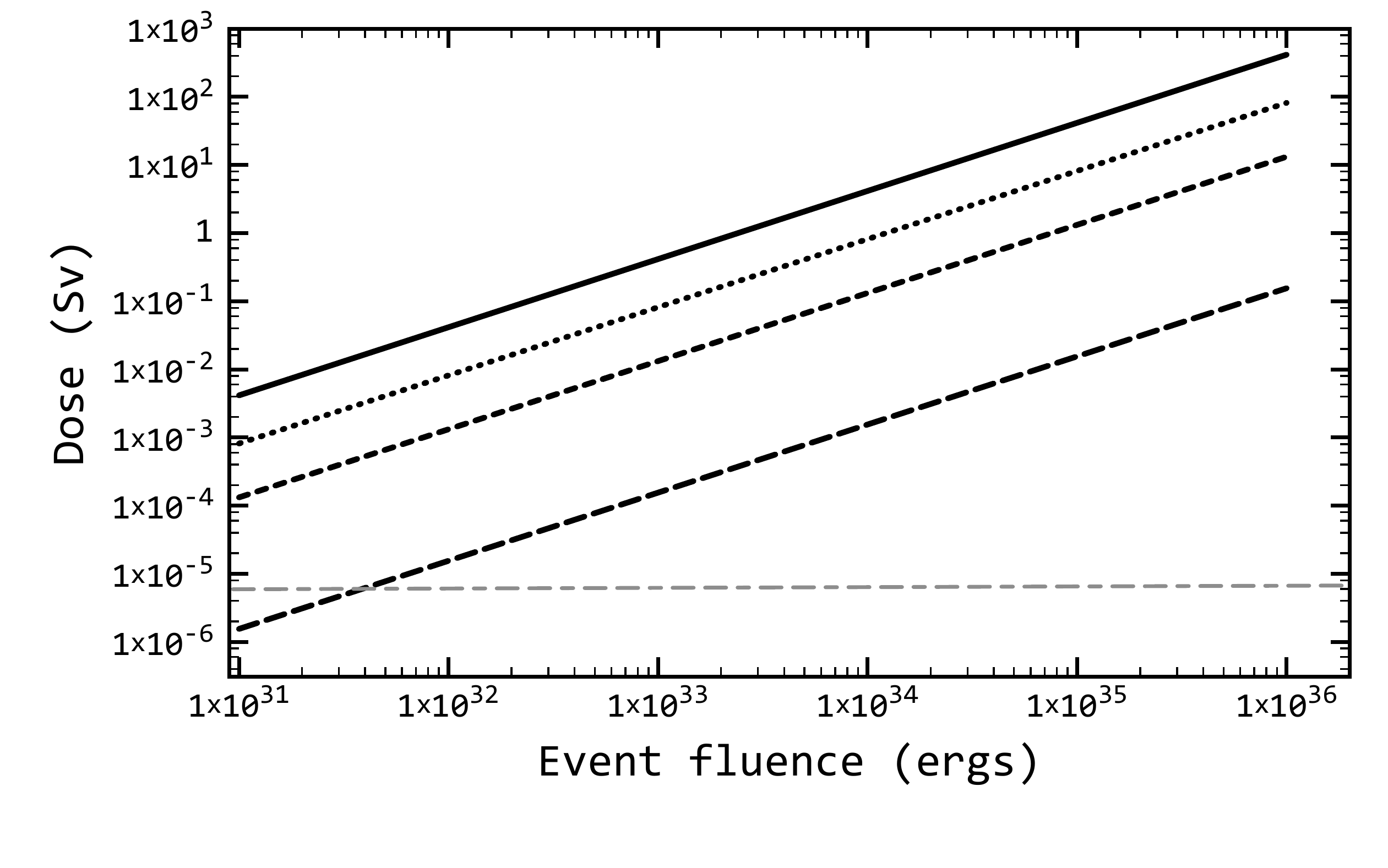}
\caption{\label{fig:i} Radiation dose from hard-spectrum SPEs (SPE56 spectrum) at depths of 100 (solid), 300 (dots), 700 (short dash), 1000 g\,cm$^{-2}$ (long dash) at for Proxima Centauri b with 1 ${\cal M}$$_{Earth}$ magnetic moment. The globally averaged daily dose from natural background radiation on Earth is shown by the horizontal dashed line for comparison.}
\end{figure}

\section{Discussion}

The results show a wide range of surface radiation doses on planets in close-in configurations with varying atmospheric column depths, magnetic moments and orbital radii. The critical importance of both the planetary magnetic field and size of the atmosphere in maintaining a low radiation dose environment has been demonstrated, even in case of extreme events such as superflares. The overall effect of radiation dose would be from the cumulative effect of multiple SPE events and depend on a combination of flare energy and frequency in a particular star system. M dwarfs are very active and have the highest flare frequency, followed by K and G-type stars like the Sun \citep{hawley2014kepler}. On M dwarfs the superflare frequency in the 10$^{34}$-10$^{35}$ erg range is about once a decade, 10$^{33}$ erg flares occur once every 100 days and 10$^{32}$ erg flares occur about once every 5 days \citep{maehara2012superflares,hawley2014kepler}. In general large superflares are more frequent by a factor of 20 on M and 6 on K-type stars compared to G-type stars\citep{shibayama2013}. The occurrence frequency of superflares can be fitted with a power law which is $\sim$ -1.6 for M stars, $\sim$-1.7 for K and much steeper $\sim$ -2.2 for G-type stars \citep{maehara2012superflares}. Statistical analysis of superflares have indicated that 10$^{34}$-10$^{35}$ ergs superflares can occur once every 800-5000 yr on a Sun-like star \citep{maehara2012superflares,shibata2003can,shibayama2013}. Young and fast-rotating stars are expected to have such flares with higher intensity and frequency consistent with dynamo theory, however, not all superflare stars fit this criteria \citep{2014A&A...567A..36W}. Some suggest that superflares could occur on stars with relatively low activity levels too \citep{candelaresi2014superflare}. More observations will give us a better picture of the rate of occurrence of these events and would help in better estimating the threat to potential planetary life in such systems.

Based on the results of radiation exposure experiments, it is seen that organisms are resistant to a wide range of radiation doses, ranging from a few Sv to $\sim$10$^5$ Sv. A radiation dose of 50 mSv and beyond is considered harmful for humans and can cause mutations and lead to carcinogenic effects. A dose of 1 Sv causes radiation sickness in humans and higher doses can cause death. Primitive microorganisms (i.e.\,E. Coli) are generally more radiation resistant than modern animals (i.e.\,mammals). Lethal radiation doses are estimated to be 5-10 Sv for humans, 6-10 Sv for small mammals, 100 Sv for goldfish, 150 Sv for bats, 100-1000 Sv for insects, 10$^4$ Sv for viruses, and 10$^4$-10$^5$ Sv for {\it Deinococcus radiodurans} \citep{harrison1996taxonomic}. Such high radiation doses on a planet can limit the type of organisms that could survive in such conditions. For example, a dose of 1000 Sv would eliminate all mammals, birds and insects and would leave the planet with more primitive lifeforms such as viruses, {\it Deinococcus radiodurans} and other extremophiles. A dose above 10$^5$ Sv would practically sterilize the planet and leave it uninhabitable to life as we know it. None of the cases considered here lead to such high radiation dose. 

Based on the simulation results, it can be concluded that for close-in exoplanets with sizable atmospheres and magnetospheres, the radiation dose contributed by stellar superflares is not enough to sterilize a planet (for life as we know it) but can result in frequent extinction level events shaping the evolution of potential ecosystems on such exoplanets. These results are especially important for exoplanets around low mass stars such as M dwarfs whose habitable zones lie in close-in configurations.

\section*{Acknowledgements}

The author thanks Jean-Mathias Grie{\ss}meier and the anonymous reviewer for their helpful comments on the manuscript, Allan Tylka for providing the table of SPE parameters and acknowledges the developers of GEANT4 (geant4.cern.ch) and ROOT (root.cern.ch). This work made use of the Extreme Science and Engineering Discovery Environment, supported by the National Science Foundation grant number ACI- 1053575.




\bibliographystyle{mnras}
\bibliography{spe-exo} 

\begin{thebibliography}{}
\makeatletter
\relax
\def\mn@urlcharsother{\let\do\@makeother \do\$\do\&\do\#\do\^\do\_\do\%\do\~}
\def\mn@doi{\begingroup\mn@urlcharsother \@ifnextchar [ {\mn@doi@}
  {\mn@doi@[]}}
\def\mn@doi@[#1]#2{\def\@tempa{#1}\ifx\@tempa\@empty \href
  {http://dx.doi.org/#2} {doi:#2}\else \href {http://dx.doi.org/#2} {#1}\fi
  \endgroup}
\def\mn@eprint#1#2{\mn@eprint@#1:#2::\@nil}
\def\mn@eprint@arXiv#1{\href {http://arxiv.org/abs/#1} {{\tt arXiv:#1}}}
\def\mn@eprint@dblp#1{\href {http://dblp.uni-trier.de/rec/bibtex/#1.xml}
  {dblp:#1}}
\def\mn@eprint@#1:#2:#3:#4\@nil{\def\@tempa {#1}\def\@tempb {#2}\def\@tempc
  {#3}\ifx \@tempc \@empty \let \@tempc \@tempb \let \@tempb \@tempa \fi \ifx
  \@tempb \@empty \def\@tempb {arXiv}\fi \@ifundefined
  {mn@eprint@\@tempb}{\@tempb:\@tempc}{\expandafter \expandafter \csname
  mn@eprint@\@tempb\endcsname \expandafter{\@tempc}}}

\bibitem[\protect\citeauthoryear{Agostinelli et~al.,}{Agostinelli
  et~al.}{2003}]{agostinelli2003geant4}
Agostinelli S.,  et~al., 2003, Nucl. Instr. Meth. Phys. Res. A, 506, 250

\bibitem[\protect\citeauthoryear{Anglada-Escud{\'e} et~al.,}{Anglada-Escud{\'e}
  et~al.}{2016}]{anglada2016terrestrial}
Anglada-Escud{\'e} G.,  et~al., 2016, Nature, 536, 437

\bibitem[\protect\citeauthoryear{Atri \& Melott}{Atri \&
  Melott}{2011}]{atri2011modeling}
Atri D.,  Melott A.~L.,  2011, Radiation Physics and Chemistry, 80, 701

\bibitem[\protect\citeauthoryear{Atri \& Melott}{Atri \&
  Melott}{2014}]{atri2014cosmic}
Atri D.,  Melott A.~L.,  2014, Astroparticle Physics, 53, 186

\bibitem[\protect\citeauthoryear{Atri, Melott  \& Thomas}{Atri
  et~al.}{2010}]{atri2010lookup}
Atri D.,  Melott A.~L.,   Thomas B.~C.,  2010, JCAP, 2010, 008

\bibitem[\protect\citeauthoryear{Atri, Hariharan  \& Grie{\ss}meier}{Atri
  et~al.}{2013}]{atri2013galactic}
Atri D.,  Hariharan B.,   Grie{\ss}meier J.-M.,  2013, Astrobiology, 13, 910

\bibitem[\protect\citeauthoryear{Candelaresi, Hillier, Maehara, Brandenburg  \&
  Shibata}{Candelaresi et~al.}{2014}]{candelaresi2014superflare}
Candelaresi S.,  Hillier A.,  Maehara H.,  Brandenburg A.,   Shibata K.,  2014,
  \apj, 792, 67

\bibitem[\protect\citeauthoryear{Dartnell}{Dartnell}{2011}]{dartnell2011ionizing}
Dartnell L.~R.,  2011, Astrobiology, 11, 551

\bibitem[\protect\citeauthoryear{Ferrari \& Szuszkiewicz}{Ferrari \&
  Szuszkiewicz}{2009}]{ferrari2009cosmic}
Ferrari F.,  Szuszkiewicz E.,  2009, Astrobiology, 9, 413

\bibitem[\protect\citeauthoryear{Grie{\ss}meier, Tabataba-Vakili, Stadelmann,
  Grenfell  \& Atri}{Grie{\ss}meier et~al.}{2015}]{griessmeier2015galactic}
Grie{\ss}meier J.-M.,  Tabataba-Vakili F.,  Stadelmann A.,  Grenfell J.,   Atri
  D.,  2015, \aap, 581, A44

\bibitem[\protect\citeauthoryear{Grie{\ss}meier, Tabataba-Vakili, Stadelmann,
  Grenfell  \& Atri}{Grie{\ss}meier et~al.}{2016}]{griessmeier2016galactic}
Grie{\ss}meier J.-M.,  Tabataba-Vakili F.,  Stadelmann A.,  Grenfell J.,   Atri
  D.,  2016, \aap, 587, A159

\bibitem[\protect\citeauthoryear{Haberle, McKay, Tyler  \& Reynolds}{Haberle
  et~al.}{1996}]{haberle1996can}
Haberle R.~M.,  McKay C.~P.,  Tyler D.,   Reynolds R.~T.,  1996, in
  Circumstellar Habitable Zones. p.~29

\bibitem[\protect\citeauthoryear{Harrison \& Anderson}{Harrison \&
  Anderson}{1996}]{harrison1996taxonomic}
Harrison F.,  Anderson S.,  1996, in Proceedings Of the Symposium: Ionizing
  Radiation, the Swedish Radiation. Protection Institute (SSI) and the Atomic
  Energy Control Board (AECB) of Canada. pp 20--24

\bibitem[\protect\citeauthoryear{Hawley, Davenport, Kowalski, Wisniewski, Hebb,
  Deitrick  \& Hilton}{Hawley et~al.}{2014}]{hawley2014kepler}
Hawley S.~L.,  Davenport J.~R.,  Kowalski A.~F.,  Wisniewski J.~P.,  Hebb L.,
  Deitrick R.,   Hilton E.~J.,  2014, \apj, 797, 121

\bibitem[\protect\citeauthoryear{Maehara et~al.,}{Maehara
  et~al.}{2012}]{maehara2012superflares}
Maehara H.,  et~al., 2012, Nature, 485, 478

\bibitem[\protect\citeauthoryear{Melott \& Thomas}{Melott \&
  Thomas}{2011}]{melott2011astrophysical}
Melott A.~L.,  Thomas B.~C.,  2011, Astrobiology, 11, 343

\bibitem[\protect\citeauthoryear{Segura, Kasting, Meadows, Cohen, Scalo, Crisp,
  Butler  \& Tinetti}{Segura et~al.}{2005}]{segura2005biosignatures}
Segura A.,  Kasting J.~F.,  Meadows V.,  Cohen M.,  Scalo J.,  Crisp D.,
  Butler R.~A.,   Tinetti G.,  2005, Astrobiology, 5, 706

\bibitem[\protect\citeauthoryear{Segura, Walkowicz, Meadows, Kasting  \&
  Hawley}{Segura et~al.}{2010}]{segura2010effect}
Segura A.,  Walkowicz L.~M.,  Meadows V.,  Kasting J.,   Hawley S.,  2010,
  Astrobiology, 10, 751

\bibitem[\protect\citeauthoryear{{Shibata} et~al.,}{{Shibata}
  et~al.}{2013}]{shibata2003can}
{Shibata} K.,  et~al., 2013, \mn@doi [PASJ] {10.1093/pasj/65.3.49}, \href
  {http://adsabs.harvard.edu/abs/2013PASJ...65...49S} {65}

\bibitem[\protect\citeauthoryear{{Shibayama} et~al.,}{{Shibayama}
  et~al.}{2013}]{shibayama2013}
{Shibayama} T.,  et~al., 2013, \mn@doi [\\apjs] {10.1088/0067-0049/209/1/5},
  \href {http://adsabs.harvard.edu/abs/2013\\apjs..209....5S} {209, 5}

\bibitem[\protect\citeauthoryear{{Smart}, {Shea}, {Spence}  \& {Kepko}}{{Smart}
  et~al.}{2006}]{smart2006}
{Smart} D.~F.,  {Shea} M.~A.,  {Spence} H.~E.,   {Kepko} L.,  2006, \mn@doi
  [AdSpR] {10.1016/j.asr.2005.09.008}, \href
  {http://adsabs.harvard.edu/abs/2006AdSpR..37.1734S} {37, 1734}

\bibitem[\protect\citeauthoryear{{Tabataba-Vakili}, {Grenfell},
  {Grie{\ss}meier}  \& {Rauer}}{{Tabataba-Vakili}
  et~al.}{2016}]{2016A&A...585A..96T}
{Tabataba-Vakili} F.,  {Grenfell} J.~L.,  {Grie{\ss}meier} J.-M.,   {Rauer} H.,
   2016, \mn@doi [\aap] {10.1051/0004-6361/201425602}, \href
  {http://adsabs.harvard.edu/abs/2016A%26A...585A..96T} {585, A96}

\bibitem[\protect\citeauthoryear{Tylka \& Dietrich}{Tylka \&
  Dietrich}{2009}]{tylka2009new}
Tylka A.~J.,  Dietrich W.~F.,  2009, in 31th International Cosmic Ray
  Conference.

\bibitem[\protect\citeauthoryear{{Vidotto}, {Fares}, {Jardine}, {Moutou}  \&
  {Donati}}{{Vidotto} et~al.}{2015}]{2015MNRAS.449.4117V}
{Vidotto} A.~A.,  {Fares} R.,  {Jardine} M.,  {Moutou} C.,   {Donati} J.-F.,
  2015, \mn@doi [MNRAS] {10.1093/mnras/stv618}, \href
  {http://adsabs.harvard.edu/abs/2015MNRAS.449.4117V} {449, 4117}

\bibitem[\protect\citeauthoryear{{Wichmann}, {Fuhrmeister}, {Wolter}  \&
  {Nagel}}{{Wichmann} et~al.}{2014}]{2014A&A...567A..36W}
{Wichmann} R.,  {Fuhrmeister} B.,  {Wolter} U.,   {Nagel} E.,  2014, \mn@doi
  [\aap] {10.1051/0004-6361/201423717}, \href
  {http://adsabs.harvard.edu/abs/2014A%26A...567A..36W} {567, A36}

\makeatother
\end{thebibliography}





\bsp	
\label{lastpage}
\end{document}